\documentclass[11pt, a4paper, sort&compress, numbers,showpacs,preprintnumbers, floatfix]{article}

\usepackage{amsmath}
\usepackage{amsfonts}
\usepackage{amssymb}
\usepackage{graphicx}
\usepackage{bm}

\def\be{\begin{equation}}

\def\ee{\end{equation}}

\def\bi{\bibitem}

\begin{document}

\title{Canonical equivalence in anisotropic models for higher order theory of gravity}

\author{Subhra Debnath\footnote{E-mail:
subhra\_ dbnth@yahoo.com}~ and Abhik Kumar Sanyal\footnote{E-mail: sanyal\_ ak@yahoo.com}}

\maketitle
\flushbottom

\begin{center}
Dept. of Physics, Jangipur College, Murshidabad, India - 742213
\end{center}

\begin{abstract}
We show that as in the case of isotropic models, the `Dirac Algorithm' and `Modified Horowitz' Formalism' lead to identical phase-space structure of the Hamiltonian for the gravitational action with curvature squared terms, in anisotropic space-time, viz, Bianchi-I, Bianchi-III and Kantowski-Sachs models too.
\end{abstract}

\section{Introduction}

Canonical formulation for higher order theories may only be possible by seeking additional degrees of freedom, which was initiated by Ostrogradski long back \cite{Os}. However, if the Hessian determinant vanishes and the Lagrangian turns out to be singular, the formalism \cite{Os} does not apply, and Dirac's constrained analysis \cite{D1, D2} appears to rescue. At sufficiently short distance ($l_P \sim 10^{-35}$ m), gravity is required to be quantized. General Theory of Relativity (GTR) being non-renormalizable, Einstein-Hilbert action is necessarily required to be supplemented by higher order curvature invariant terms \cite{Stelle}. Further, with the advent of superstring, heterotic string \cite{S1, S2} and supergravity \cite{S3, S4} theories, it is now quite apparent that under weak field approximation, these theories primarily reduce to effective actions containing higher-order curvature invariant terms in $4$-dimensions. In the absence of a complete quantum theory of gravity, quantum cosmology renders certain insights near Planck's era. Hence, to study early universe, canonical formulation of gravity with higher-order curvature invariant terms is necessitated.\\

While GTR may be expressed only in terms of the basic variables $h_{ij}$ - the induced three-metric, canonical formulation of gravity with higher-order curvature invariant terms, requires additional degree of freedom, as mentioned. The extrinsic curvature tensor - $K_{ij}$ plays the role of these additional basic variables. For canonical formulation of higher-order theory of gravity, following Boulware's formalism \cite{Boul}, Horowitz' proposed a technique that bypasses Dirac's constraint analysis elegantly even if the Hessian determinant ceases to exist \cite{Horo}, and may be applied in general for actions with arbitrary curvature invariant terms. The main essence of Horowitz' formalism is primarily to express the action in terms of the basic variable $h_{ij}$, and to find auxiliary variables as the derivative of the `action' with respect to the highest derivative appearing in it. The auxiliary variables are then judiciously substituted, so that the action turns out to be canonical. The Hamiltonian is then expressed with respect to the basic variables and the auxiliary variables. Finally, the auxiliary variables are replaced by the basic variables $K_{ij}$ - the extrinsic curvature tensor, following canonical transformations, and in the process, one ends up with the phase-space structure of the Hamiltonian.\\

Pollock \cite{Pol} first realized that the Horowitz' formalism suffers from the problem of dealing with systems which do not even contain higher-order terms and hence ends up with a wrong quantum dynamics. It has additionally been shown that Horowitz' formalism suffers from certain ir-repairable pathologies, in the sense that it removes additional total derivative terms form the action, which do not match with those found under variation of the action \cite{Ab1, Ab2}. Later it was observed that Ostrogradski's, Dirac's and Horowitz' formalisms lead to the same Hamiltonian \cite{Ab3, Ab4}, while Modified Horowitz' Formalism (MHF), in which one takes care of the divergent terms appearing in the action, prior to the introduction of the auxiliary variable and follow Horowitz' formalism thereafter, leads to a different Hamiltonian, with some beautiful features \cite{Ab1, Ab5, Ab6, Ab7}. These features include particularly i) the de-Sitter form of expansion upon extremization of the effective potential, ii) fixation of the operator ordering parameter from physical ground (establishing unitarity of the Hamiltonian, and rendering a straight forward quantum mechanical probabilistic interpretation), iii) admittance of oscillatory behaviour about classical de-Sitter solution of the semiclassical wavefunction etc. \cite{Ab2, Ab8}. Recently, it has further been observed that even Dirac's formalism after taking care of the total derivative terms appearing in the action leads to the same phase-space structure of the Hamiltonian as found following MHF \cite{Ab9}. Indeed the different Hamiltonians are related under suitable canonical transformation in the classical domain. However, due to non-linearity, no quantum canonical transformation can relate the two \cite{Ab9}. Clearly canonically equivalent Hamiltonians at the classical domain does not justify equivalence at the quantum domain. Since, MHF and Dirac's constrained analysis (after taking care of divergent terms) yield identical Hamiltonian and have shown no pathology so far, should be treated as the correct formalisms towards canonical formulation of higher-order theory of gravity.\\

Nonetheless all these results appeared in isotropic and homogeneous mini-superspace models being guided by cosmological principle. It is therefore worth to see if the same (identical Hamiltonian structure following MHF and Dirac's formalism after taking care of the divergent terms) is true upon relaxing the cosmological principle by taking anisotropic models into account. In the present manuscript, we therefore consider homogeneous-anisotropic-axially symmetric, Bianchi-I, Bianchi-III and Kantowski-Sachs minisuperspace models, and study the phase-space structure of a generalised action containing curvature squared terms ($R^2,~R_{\mu\nu}^2$), with arbitrary functional dependence on a scalar field, following both the Dirac's formalism (after taking care of the divergent terms) and MHF. \\

A qualitative study of Kantowski-Sachs (KS) \cite{KS} cosmological models had been performed long back \cite{Web1, Web2}. It was found that these models are spatially homogeneous, non-rotating, have shear, and do not belong to the Bianchi classes. Weber \cite{Web1} particularly investigated such models with a cosmological constant and found that there exist models evolving towards the de Sitter universe. Such cosmological models are of particular interest due to the possible existence of a GUT phase transition producing a vacuum-dominated inflationary era in the very early history of the universe \cite{Linde}.\\

In the following section, we write the action, field equations and present an analytic de-Sitter solution in vacuum. In section 3, we first perform Dirac's constrained analysis to explore the phase-space structure of the Hamiltonian. We also compute the same following MHF, thereafter. We conclude in section 4.

\section{Action, Field equations and Classical de-Sitter solution:}

Our starting point is the following general fourth-order gravitational action \cite{Barth}

\be\label{A} A=\int\Bigg[\alpha(\phi)R + \beta(\phi)R^2 + \gamma(\phi)R_{\mu\nu}R^{\mu\nu} - \frac{1}{2}\phi_{,\mu}\phi^{,\mu} - V(\phi) \Bigg]\sqrt{-g}d^4x, \ee
in which all the coupling parameters are arbitrary functions of the scalar field $\phi$. Under metric variation, one obtains the following field equation,

\be\begin{split} \alpha G_{\mu\nu} &+ g_{\mu\nu}\Box\alpha - \alpha_{;\mu\nu} + \beta\Big(2RR_{\mu\nu} - \frac{1}{2}g_{\mu\nu}R^2\Big)  + 2g_{\mu\nu}\Box (\beta R) - 2(\beta R)_{;{\mu\nu}}\\
& + \gamma\Big(2R^{\alpha\beta}R_{\alpha\mu\beta\nu} - \frac{1}{2}g_{\mu\nu}R_{\alpha\beta}R^{\alpha\beta}\Big) + \frac{1}{2}g_{\mu\nu}\Box (\gamma R) + \Box (\gamma R_{\mu\nu}) - (\gamma R)_{;{\mu\nu}} = \frac{T_{\mu\nu}}{2},\end{split}\ee
where, $G_{\mu\nu} = R_{\mu\nu} - \frac{1}{2}g_{\mu\nu}R$ and $T_{\mu\nu} = \nabla_\mu\phi\nabla_\nu\phi - \frac{1}{2}g_{\mu\nu}\nabla_\lambda\phi\nabla^\lambda\phi - g_{\mu\nu}V(\phi)$ are the Einstein tensor and the energy-momentum tensor, respectively. Variation of the action \eqref{A} with respect to $\phi$ yields the following generalized Klein-Gordon equation,

\be \Box\phi - \alpha'R - \beta' R^2 - \gamma'R_{\mu\nu}R^{\mu\nu} - V' = 0.\ee
While turning our attention towards minisuperspace model, we express the homogeneous and anisotropic Kantowski-Sachs, axially symmetric Bianchi-I and Bianchi-III minisuperspace models altogether, in the following form,

\be\label{aniso}\begin{split} & ds^2 = -dt^2 + A(t)^2dr^2 + B(t)^2(d\theta^2 + f^2_k(\theta) d\phi^2),  ~~\mathrm{where},\\&
f_k = \sin\theta, ~~~~~~~~\mathrm{Kantowski-Sachs ~metric~ with~ positive~ spatial~ curvature}, k = 1,\\&
f_k = \theta,~~~~~~~~~~~~\mathrm{Bianchi-I~ metric ~with~ zero~ spatial~ curvature}, k = 0, \\&
f_k = \sin h\theta, ~~~~~~\mathrm{Bianchi-III ~metric ~with ~negative ~spatial~ curvature}, k=-1, \end{split}\ee
$k$ being the curvature index of the 2-dimensional surface $d\theta^2 + f^2_k (\theta)d\phi^2$. Correspondingly, the Ricci scalar and the Ricci tensor are given by,

\be \begin{split} &R = \frac{2}{N^2} \Bigg(\frac{\ddot A}{A} + 2 \frac{\ddot B}{B} + 2\frac{\dot A}{A}\frac{\dot B}{B} + \frac{\dot B^2}{B^2} + \frac{k N^2}{B^2} - \frac{\dot A}{A}\frac{\dot N}{N} - 2\frac{\dot B}{B}\frac{\dot N}{N}\Bigg),\\&
R_{\mu\nu}R^{\mu\nu} = \frac{1}{N^4} \Bigg[\Bigg(\frac{\ddot A}{A} + 2\frac{\ddot B}{B} - \frac{\dot A}{A}\frac{\dot N}{N} - 2\frac{\dot B}{B}\frac{\dot N}{N}\Bigg)^2 + \Bigg(\frac{\ddot A}{A} + 2 \frac{\dot A}{A}\frac{\dot B}{B} - \frac{\dot A}{A}\frac{\dot N}{N}\Bigg)^2 \\&
\hspace{1.0 in}+ 2\Bigg(\frac{\ddot B}{B} + \frac{\dot A}{A}\frac{\dot B}{B} + \frac{\dot B^2}{B^2} + \frac{k N^2}{B^2} - \frac{\dot B}{B}\frac{\dot N}{N}\Bigg)^2\Bigg].\end{split}\ee
The field equations, viz. the $A$, $B$, $\phi$ variation equations and the ($^0_0$) equation of Einstein respectively are the following,

\be\label{z}\begin{split} &2\alpha\Bigg(2\frac{\ddot B}{B} + \frac{\dot B^2}{B^2} + \frac{k}{B^2}\Bigg) + 4\beta\Bigg(2\frac{\ddddot A}{A} - 4\frac{\dot A\dddot A}{A^2} + 8\frac{\dot B\dddot A}{AB} + 4\frac{\dot A\dddot B}{AB} + 4\frac{\dot B\dddot B}{B^2} + 4\frac{\ddddot B}{B} - 3\frac{\ddot A^2}{A^2} + 4\frac{\dot A^2\ddot A}{A^3}\\
& + 4\frac{\ddot B^2}{B^2} - 8\frac{\dot A^2\ddot B}{A^2B} - 8\frac{\dot B^2\ddot B}{B^3} - 20\frac{\dot A\dot B\ddot A}{A^2B}- 4\frac{\dot A\dot B\ddot B}{AB^2} + 8\frac{\ddot A\ddot B}{AB} + 8\frac{\dot A^3\dot B}{A^3B} - 4\frac{\dot A^2\dot B^2}{A^2B^2} + 5\frac{\dot B^4}{B^4} + 6\frac{k\dot B^2}{B^4} + \frac{k^2}{B^4}\Bigg)\\
& + 2\gamma\Bigg(2\frac{\ddddot A}{A} - 4\frac{\dot A\dddot A}{A^2} + 8\frac{\dot B\dddot A}{AB} + 2\frac{\dot A\dddot B}{AB} + 2\frac{\dot B\dddot B}{B^2} + 2\frac{\ddddot B}{B} - 3\frac{\ddot A^2}{A^2} + 4\frac{\dot A^2\ddot A}{A^3} + 3\frac{\ddot B^2}{B^2} - 4\frac{\dot A^2\ddot B}{A^2B} - 4\frac{\dot B^2\ddot B}{B^3}\\
& - 14\frac{\dot A\dot B\ddot A}{A^2B} - 6\frac{\dot A\dot B\ddot B}{AB^2} + 6\frac{\ddot A\ddot B}{AB} + 4\frac{\dot A^3\dot B}{A^3B} - \frac{\dot A^2\dot B^2}{A^2B^2} + 3\frac{\dot B^4}{B^4} + 4\frac{k\dot B^2}{B^4} + \frac{k^2}{B^4}\Bigg) + 4\alpha'\dot\phi\frac{\dot B}{B} + 16\beta'\dot\phi\Bigg(\frac{\dddot A}{A}\\
& + 2\frac{\dddot B}{B} - \frac{\dot A\ddot A}{A^2} + 3\frac{\dot B\ddot A}{AB} + 2\frac{\dot A\ddot B}{AB} + 2\frac{\dot B\ddot B}{B^2} - \frac{k\dot B}{B^3} - 2\frac{\dot A^2\dot B}{A^2B} - \frac{\dot B^3}{B^3}\Bigg) + 4\gamma'\dot\phi\Bigg(2\frac{\dddot A}{A} + 2\frac{\dddot B}{B} - 2\frac{\dot A\ddot A}{A^2} + 5\frac{\dot B\ddot A}{AB}\\
& + 2\frac{\dot A\ddot B}{AB} + \frac{\dot B\ddot B}{B^2} - \frac{k\dot B}{B^3} - 2\frac{\dot A^2\dot B}{A^2B} - \frac{\dot A\dot B^2}{AB^2} - \frac{\dot B^3}{B^3}\Bigg) + 8\big(\beta'\ddot\phi + \dot\phi^2\beta''\big)\Bigg(\frac{\ddot A}{A} + 2\frac{\ddot B}{B} + 2\frac{\dot A\dot B}{AB} + \frac{\dot B^2}{B^2} + \frac{k}{B^2}\Bigg)\\
& + 4\big(\gamma'\ddot\phi + \dot\phi^2\gamma''\big)\Bigg(\frac{\ddot A}{A} + \frac{\ddot B}{B} + \frac{\dot A\dot B}{AB}\Bigg) + 2\big(\alpha'\ddot\phi + \dot\phi^2\alpha''\big) + \Bigg(\frac{\dot\phi^2}{2} - V\Bigg)= 0\end{split}\ee

\be\label{y}\begin{split} &2\alpha\Bigg(\frac{\ddot A}{A} + \frac{\ddot B}{B} + \frac{\dot A\dot B}{AB}\Bigg) + 4\beta\Bigg(2\frac{\ddddot A}{A} - 2\frac{\dot A\dddot A}{A^2} + 6\frac{\dot B\dddot A}{AB} + 8\frac{\dot A\dddot B}{AB} + 4\frac{\ddddot B}{B} - \frac{\ddot A^2}{A^2} + 2\frac{\dot A^2\ddot A}{A^3} - 4\frac{\dot B^2\ddot A}{AB^2}\\
& - 4\frac{\dot A^2\ddot B}{A^2B} - 14\frac{\dot B^2\ddot B}{B^3} - 8\frac{\dot A\dot B\ddot A}{A^2B} - 8\frac{\dot A\dot B\ddot B}{AB^2} + 10\frac{\ddot A\ddot B}{AB} - 6\frac{k\ddot B}{B^3} + 4\frac{\dot A^3\dot B}{A^3B} - 2\frac{\dot A\dot B^3}{AB^3} + 7\frac{\dot B^4}{B^4} - 6\frac{k\dot A\dot B}{AB^3}\\
& + 6\frac{k\dot B^2}{B^4} - \frac{k^2}{B^4}\Bigg) + 2\gamma\Bigg(\frac{\ddddot A}{A} - \frac{\dot A\dddot A}{A^2} + 3\frac{\dot B\dddot A}{AB} + 6\frac{\dot A\dddot B}{AB} + 3\frac{\ddddot B}{B} + \frac{\dot A^2\ddot A}{A^3} - 2\frac{\dot B^2\ddot A}{AB^2} - 3\frac{\dot A^2\ddot B}{A^2B}\\
& - 10\frac{\dot B^2\ddot B}{B^3} - 6\frac{\dot A\dot B\ddot A}{A^2B} - 4\frac{\dot A\dot B\ddot B}{AB^2} + 6\frac{\ddot A\ddot B}{AB} - 4\frac{k\ddot B}{B^3} + 3\frac{\dot A^3\dot B}{A^3B} - 2\frac{\dot A\dot B^3}{AB^3} + 5\frac{\dot B^4}{B^4} - 4\frac{k\dot A\dot B}{AB^3} + 4\frac{k\dot B^2}{B^4}\\
& - \frac{k^2}{B^4}\Bigg) + 2\alpha'\dot\phi\Bigg(\frac{\dot A}{A} + \frac{\dot B}{B}\Bigg) + 8\beta'\dot\phi\Bigg(2\frac{\dddot A}{A} + 4\frac{\dddot B}{B} - \frac{\dot A\ddot A}{A^2} + 5\frac{\dot B\ddot A}{AB} + 6\frac{\dot A\ddot B}{AB} + 2\frac{\dot B\ddot B}{B^2} + \frac{k\dot A}{AB^2} - 3\frac{k\dot B}{B^3}\\
& - 2\frac{\dot A^2\dot B}{A^2B} - \frac{\dot A\dot B^2}{AB^2} - 3\frac{\dot B^3}{B^3}\Bigg) + 2\gamma'\dot\phi\Bigg(2\frac{\dddot A}{A} + 6\frac{\dddot B}{B} - \frac{\dot A\ddot A}{A^2} + 4\frac{\dot B\ddot A}{AB} + 7\frac{\dot A\ddot B}{AB} + 2\frac{\dot B\ddot B}{B^2} + \frac{k\dot A}{AB^2} - 4\frac{k\dot B}{B^3}\\
& - 3\frac{\dot A^2\dot B}{A^2B} - \frac{\dot A\dot B^2}{AB^2} - 4\frac{\dot B^3}{B^3}\Bigg) + 2\big(\alpha'\ddot\phi + \dot\phi^2\alpha''\big) + 8\big(\beta'\ddot\phi + \dot\phi^2\beta''\big)\Bigg(\frac{\ddot A}{A} + 2\frac{\ddot B}{B} + 2\frac{\dot A\dot B}{AB} + \frac{\dot B^2}{B^2} + \frac{k}{B^2}\Bigg)\\
& + 2\big(\gamma'\ddot\phi + \dot\phi^2\gamma''\big)\Bigg(\frac{\ddot A}{A} + \frac{\ddot B}{B} + \frac{\dot A\dot B}{AB}\Bigg) + \Bigg(\frac{\dot\phi^2}{2} - V\Bigg)= 0\end{split}\ee

\be\label{phi}\begin{split} &\ddot\phi + \Bigg(\frac{\dot A}{A} + 2\frac{\dot B}{B}\Bigg)\dot\phi + V' - 2\alpha'\Bigg(\frac{\ddot A}{A} + 2\frac{\ddot B}{B} + 2\frac{\dot A\dot B}{AB} + \frac{\dot B^2}{B^2} + \frac{k}{B^2}\Bigg) - 4\beta'\Bigg(\frac{\ddot A^2}{A^2} + 4\frac{\dot A\dot B\ddot A}{A^2B}\\
& + 2\frac{\dot B^2\ddot A}{AB^2} + 2\frac{k\ddot A}{AB^2} + 4\frac{\ddot B^2}{B^2} + 8\frac{\dot A\dot B\ddot B}{AB^2} + 4\frac{\dot B^2\ddot B}{B^3} + 4\frac{k\ddot B}{B^3} + 4\frac{\ddot A\ddot B}{AB} + 4\frac{k\dot A\dot B}{AB^3} + 2\frac{k\dot B^2}{B^4} + 4\frac{\dot A^2\dot B^2}{A^2B^2}\\
& + 4\frac{\dot A\dot B^3}{AB^3} + \frac{\dot B^4}{B^4} + \frac{k^2}{B^4}\Bigg) - 2\gamma'\Bigg(\frac{\ddot A^2}{A^2} + 2\frac{\dot A\dot B\ddot A}{A^2B} + 3\frac{\ddot B^2}{B^2} + 2\frac{\dot A\dot B\ddot B}{AB^2} + 2\frac{\dot B^2\ddot B}{B^3} + 2\frac{k\ddot B}{B^3} + 2\frac{\ddot A\ddot B}{AB}\\
& + 2\frac{k\dot A\dot B}{AB^3} + 2\frac{k\dot B^2}{B^4} + 3\frac{\dot A^2\dot B^2}{A^2B^2} + 2\frac{\dot A\dot B^3}{AB^3} + \frac{\dot B^4}{B^4} + \frac{k^2}{B^4}\Bigg) = 0\end{split}\ee

\be\label{00}\begin{split} &2\alpha\Bigg(\frac{\dot B^2}{B^2} + 2\frac{\dot A\dot B}{AB} + \frac{k}{B^2}\Bigg) + 4\beta\Bigg(2\frac{\dot A\dddot A}{A^2} + 4\frac{\dot B\dddot A}{AB} + 4\frac{\dot A\dddot B}{AB} + 8\frac{\dot B\dddot B}{B^2} - \frac{\ddot A^2}{A^2} - 2\frac{\dot A^2\ddot A}{A^3} + 8\frac{\dot B^2\ddot A}{AB^2} - 4\frac{\ddot B^2}{B^2}\\
& + 4\frac{\dot A^2\ddot B}{A^2B} + 8\frac{\dot A\dot B\ddot B}{AB^2} - 4\frac{\ddot A\ddot B}{AB} - 4\frac{\dot A^3\dot B}{A^3B} - 8\frac{\dot A^2\dot B^2}{A^2B^2} - 8\frac{\dot A\dot B^3}{AB^3} - 7\frac{\dot B^4}{B^4} - 6\frac{k\dot B^2}{B^4} + \frac{k^2}{B^4}\Bigg) + 2\gamma\Bigg(2\frac{\dot A\dddot A}{A^2}\\
& + 2\frac{\dot B\dddot A}{AB} + 2\frac{\dot A\dddot B}{AB} + 6\frac{\dot B\dddot B}{B^2} - \frac{\ddot A^2}{A^2} - 2\frac{\dot A^2\ddot A}{A^3} + 2\frac{\dot A\dot B\ddot A}{A^2B} - 3\frac{\ddot B^2}{B^2} + 2\frac{\dot A^2\ddot B}{A^2B} + 6\frac{\dot A\dot B\ddot B}{AB^2} - 2\frac{\ddot A\ddot B}{AB} - 2\frac{\dot A^3\dot B}{A^3B}\\
& - 7\frac{\dot A^2\dot B^2}{A^2B^2} - 4\frac{\dot A\dot B^3}{AB^3} - 5\frac{\dot B^4}{B^4} - 4\frac{k\dot B^2}{B^4} + \frac{k^2}{B^4}\Bigg) + 2\alpha'\dot\phi\Bigg(\frac{\dot A}{A} + 2\frac{\dot B}{B}\Bigg) + 8\beta'\dot\phi\Bigg(\frac{\dot A\ddot A}{A^2} + 2\frac{\dot B\ddot A}{AB} + 2\frac{\dot A\ddot B}{AB}\\
& + 4\frac{\dot B\ddot B}{B^2} + \frac{k\dot A}{AB^2} + 2\frac{k\dot B}{B^3} + 2\frac{\dot A^2\dot B}{A^2B} + 5\frac{\dot A\dot B^2}{AB^2} + 2\frac{\dot B^3}{B^3}\Bigg) + 4\gamma'\dot\phi\Bigg(\frac{\dot A\ddot A}{A^2} + \frac{\dot B\ddot A}{AB} + \frac{\dot A\ddot B}{AB} + 3\frac{\dot B\ddot B}{B^2} + \frac{k\dot B}{B^3}\\
& + \frac{\dot A^2\dot B}{A^2B} + \frac{\dot A\dot B^2}{AB^2} + \frac{\dot B^3}{B^3}\Bigg) - \Bigg(\frac{\dot\phi^2}{2} + V\Bigg)= 0\end{split}\ee
The field equations, (\ref{z}), (\ref{y}), (\ref{phi}) and (\ref{00}) admit the following de-Sitter solution,
\be\label{Sol}\begin{split}& A = A_0 e^{\frac{\sqrt 3 + 2c}{\sqrt 3}\lambda t},~~B = B_0 e^{\frac{\sqrt 3 - c}{\sqrt 3}\lambda t}, ~\mathrm{and}, ~\phi = \phi_0 e^{-\lambda t},\\&
\mathrm{so~that~the ~three ~volume ~is:}~~ AB^2 = A_0B_0^2 e^{3\lambda t},\\&
\mathrm{and~the~average~expansion~scale~factor~is:}~~a(t) = (AB^2)^{1\over 3}=(A_0B_0^2)^{1\over 3}e^{\lambda t},\\&
\mathrm{under~ the~ choice}, ~ ~k = 0, ~V = V_0\phi^2, ~\alpha = \alpha_0\phi^2, ~\beta = \beta_0\phi^2, ~\mathrm{and} ~\gamma = \gamma_0\phi^2, \mathrm{and~setting}\\& \alpha_0 = \frac{1}{c^2 + 6}\Big(\frac{V_0}{\lambda^2} - \frac{5c^2 + 18}{2c^2 + 9}\Big), ~~\beta_0 = -\frac{1}{4\lambda^2(c^2 + 6)^2}\Big(\frac{V_0}{\lambda^2} + \frac{c^2 + 2}{4}\Big) ~\mathrm{and},~ \gamma_0 = \frac{1}{8\lambda^2(2c^2 + 9)},
\end{split}\ee
where, $A_0, B_0, \phi_0, V_0, \alpha_0, \beta_0, \gamma_0, \lambda, c$ are constants. Clearly, the solution is for Bianchi-I universe having $k = 0$. In view of the above solution, the usual definitions of the expansion scalar and the shear scalar lead to,
\be \begin{split}&\theta = {v^{\mu}}_{;\mu} = {\dot A\over A} + 2{\dot B\over B} = 3\lambda;\hspace{0.4 cm} \sigma^2 = {1\over 2}\sigma_{\mu\nu}\sigma^{\mu\nu}= {1\over 3}\left[{\dot A\over A} - {\dot B\over B}\right]^2 =c^2 \lambda^2,\\&
\mathrm{where,}~~\sigma_{\mu\nu} = v_{(\mu;\nu)} + {1\over 2}(v_{\mu;\alpha}v^{\alpha}v_{\nu}+v_{\nu;\alpha}v^{\alpha}v_{\mu}) -{1\over \theta}(g_{\mu\nu} + v_\mu v_\nu),\end{split}\ee
where, $v_\mu$ represents the four-velocity, so that $v_\mu v^\mu = -1$. Note that during inflation, under slow-roll condition $\lambda = \mathrm{H}$, where, $\mathrm{H}$ is the slowly varying Hubble parameter, and thus, the inflation isotropizes the universe. However, inflation is not our present concern.

\section{Canonical formulation:}
As mentioned in the introduction, the Dirac's algorithm (after taking care of the divergent terms appearing in the action) and MHF lead to the identical phase-space structure of the Hamiltonian, in isotropic and homogeneous space-time. In this section, we construct the Hamiltonian corresponding to the action \eqref{A} in the anisotropic minisuperspace models under consideration \eqref{aniso} applying Dirac's algorithm in the following subsection, and thereafter following MHF in subsection (3.2).

\subsection{Dirac formalism}
In terms of basic variables, $h_{11} = A^2 \delta_{11} = z\delta_{11}$, $h_{22} = B^2\delta_{22} = y\delta_{22}$ and $h_{33}= B^2\delta_{33} = y \delta_{33}$, associated with the metric \eqref{aniso}, the action (\ref{A}) may be expressed as,

\be\label{A1}\begin{split}
A = \int&\Bigg[\frac{\alpha(\phi)}{N}\Bigg(\frac{\ddot z}{z} + 2\frac{\ddot y}{y} - \frac{\dot z^2}{2z^2} + \frac{\dot y \dot z}{yz} - \frac{\dot N\dot z}{Nz} - \frac{\dot y^2}{2y^2} - 2\frac{\dot N\dot y}{Ny} + 2\frac{kN^2}{y}\Bigg) + \frac{\beta(\phi)}{N^3}\Bigg(\frac{\ddot z^2}{z^2} + 4\frac{ \ddot y \ddot z}{y z} -\frac{\dot z^2 \ddot z}{z^3} ~+ \\
& 2\frac{\dot y\dot z\ddot z}{yz^2} - 2\frac{\dot N\dot z\ddot z}{Nz^2} - \frac{\dot y^2\ddot z}{y^2z} - 4\frac{\dot N\dot y\ddot z}{Nyz} + 4\frac{kN^2\ddot z}{yz} + 4\frac{\ddot y^2}{y^2} - 2\frac{\dot z^2\ddot y}{yz^2} + 4\frac{\dot y\dot z\ddot y}{y^2z} - 4\frac{\dot N\dot z\ddot y}{Nyz} - 2\frac{\dot y^2\ddot y}{y^3} - 8\frac{\dot N\dot y\ddot y}{Ny^2}\\
& + 8\frac{kN^2\ddot y}{y^2} + \frac{\dot z^4}{4z^4} - \frac{\dot y \dot z^3}{yz^3} + \frac{\dot N\dot z^3}{Nz^3} + \frac{3\dot y^2 \dot z^2}{2y^2 z^2} + \frac{\dot N^2\dot z^2}{N^2z^2} - 2\frac{kN^2\dot z^2}{yz^2} - \frac{\dot y^3\dot z}{y^3z} - 3\frac{\dot N\dot y^2\dot z}{Ny^2z} + 4\frac{\dot N^2\dot y\dot z}{N^2yz}\\
& + 4\frac{kN^2\dot y\dot z}{y^2z} - 4\frac{kN\dot N\dot z}{yz} + \frac{\dot y^4}{4y^4} + 2\frac{\dot N\dot y^3}{Ny^3} + 4\frac{\dot N^2\dot y^2}{N^2y^2} - 2\frac{kN^2\dot y^2}{y^3} - 8\frac{k N\dot N\dot y}{y^2} + 4\frac{k^2N^4}{y^2}\Bigg) ~+\\
& \frac{\gamma(\phi)}{N^3}\Bigg(\frac{\ddot z^2}{2z^2} + \frac{ \ddot y \ddot z}{y z} - \frac{\dot z^2 \ddot z}{2z^3} + \frac{\dot y\dot z\ddot z}{2yz^2} - \frac{\dot N\dot z\ddot z}{Nz^2} - \frac{\dot y^2\ddot z}{2y^2z} - \frac{\dot N\dot y\ddot z}{Nyz} + \frac{3\ddot y^2}{2y^2} - \frac{\dot z^2\ddot y}{2yz^2} + \frac{\dot y\dot z\ddot y}{2y^2z} - \frac{\dot N\dot z\ddot y}{Nyz}\\
& - \frac{\dot y^2\ddot y}{y^3} - 3\frac{\dot N\dot y\ddot y}{Ny^2} + 2\frac{kN^2\ddot y}{y^2} + \frac{\dot z^4}{8z^4} - \frac{\dot y \dot z^3}{4yz^3} + \frac{\dot N\dot z^3}{2Nz^3} + \frac{5\dot y^2 \dot z^2}{8y^2 z^2} + \frac{\dot N^2\dot z^2}{2N^2z^2} + \frac{\dot N^2\dot y\dot z}{N^2yz} + \frac{kN^2\dot y\dot z}{y^2z}\\
& + \frac{\dot y^4}{4y^4} + \frac{\dot N\dot y^3}{Ny^3} + \frac{3\dot N^2\dot y^2}{2N^2y^2} - 2\frac{k N\dot N\dot y}{y^2} + 2\frac{k^2N^4}{y^2}\Bigg) + \Bigg(\frac{\dot\phi^2}{2N} - NV(\phi)\Bigg) \Bigg]y\sqrt{z}dt.\end{split}\ee
The primary observation is: despite being a Lagrange multiplier, the lapse function $N$ appears in the action with its time derivative, unlike GTR. This uncanny behaviour of the lapse function appears to treat it as a dynamical variable, desisting to establish diffeomorphic invariance, $H = N\mathcal{H}$. However, one can easily compute the Hessian determinant to be sure that it vanishes, making the action degenerate. Thus to proceed with constraint analysis, let us first integrate the action (\ref{A1}) by parts as already mentioned, and remove the following total derivative terms,

\be \Bigg[\frac{\alpha(\phi)}{N}\Bigg(\frac{\dot z}{z} + 2\frac{\dot y}{y}\Bigg) + \frac{\beta(\phi)}{N^3}\Bigg(-\frac{\dot z^3}{3z^3} + 4\frac{kN^2\dot z}{yz} - \frac{2\dot y^3}{3y^3} + 8\frac{kN^2\dot y}{y^2}\Bigg) + \frac{\gamma(\phi)}{N^3}\Bigg(-\frac{\dot z^3}{6z^3} - \frac{\dot y^3}{3y^3} + 2\frac{kN^2\dot y}{y^2}\Bigg)\Bigg]y\sqrt{z}\ee
so that the action (\ref{A1}) reads as,

\be\label{A2}\begin{split}
A = \int&\Bigg[\frac{\alpha}{N}\Bigg( - \frac{\dot y \dot z}{yz}  - \frac{\dot y^2}{2y^2}  + 2\frac{kN^2}{y}\Bigg) - \frac{\alpha'\dot\phi}{N}\Bigg(\frac{\dot z}{z} + 2\frac{\dot y}{y}\Bigg) + \frac{\beta}{N^3}\Bigg(\frac{\ddot z^2}{z^2} + 4\frac{ \ddot y \ddot z}{y z} + 2\frac{\dot y\dot z\ddot z}{yz^2} - 2\frac{\dot N\dot z\ddot z}{Nz^2} - \frac{\dot y^2\ddot z}{y^2z}\\
& - 4\frac{\dot N\dot y\ddot z}{Nyz}  + 4\frac{\ddot y^2}{y^2} - 2\frac{\dot z^2\ddot y}{yz^2} + 4\frac{\dot y\dot z\ddot y}{y^2z} - 4\frac{\dot N\dot z\ddot y}{Nyz} - 8\frac{\dot N\dot y\ddot y}{Ny^2}  - \frac{7\dot z^4}{12z^4} - \frac{2\dot y \dot z^3}{3yz^3}  + \frac{3\dot y^2 \dot z^2}{2y^2 z^2} + \frac{\dot N^2\dot z^2}{N^2z^2}\\
&  - \frac{2\dot y^3\dot z}{3y^3z} - 3\frac{\dot N\dot y^2\dot z}{Ny^2z} + 4\frac{\dot N^2\dot y\dot z}{N^2yz} - \frac{13\dot y^4}{12y^4} + 4\frac{\dot N^2\dot y^2}{N^2y^2} + 6\frac{k N^2\dot y^2}{y^3} + 4\frac{k^2N^4}{y^2}\Bigg) + \frac{\beta'\dot\phi}{N^3}\Bigg(\frac{\dot z^3}{3z^3} - 4\frac{kN^2\dot z}{yz}\\
& + \frac{2\dot y^3}{3y^3} - 8\frac{kN^2\dot y}{y^2}\Bigg) + \frac{\gamma}{N^3}\Bigg(\frac{\ddot z^2}{2z^2} + \frac{ \ddot y \ddot z}{y z} + \frac{\dot y\dot z\ddot z}{2yz^2} - \frac{\dot N\dot z\ddot z}{Nz^2} - \frac{\dot y^2\ddot z}{2y^2z} - \frac{\dot N\dot y\ddot z}{Nyz} + \frac{3\ddot y^2}{2y^2} - \frac{\dot z^2\ddot y}{2yz^2} + \frac{\dot y\dot z\ddot y}{2y^2z}\\
& - \frac{\dot N\dot z\ddot y}{Nyz} - 3\frac{\dot N\dot y\ddot y}{Ny^2} - \frac{7\dot z^4}{24z^4} - \frac{\dot y \dot z^3}{12yz^3} + \frac{5\dot y^2 \dot z^2}{8y^2 z^2} + \frac{\dot N^2\dot z^2}{2N^2z^2} + \frac{\dot y^3\dot z}{6y^3z} + \frac{\dot N^2\dot y\dot z}{N^2yz} - \frac{5\dot y^4}{12y^4} + \frac{3\dot N^2\dot y^2}{2N^2y^2}\\
& + 2\frac{k N^2\dot y^2}{y^3} + 2\frac{k^2N^4}{y^2}\Bigg) + \frac{\gamma'\dot\phi}{N^3}\Bigg(\frac{\dot z^3}{6z^3} + \frac{\dot y^3}{3y^3} - 2\frac{kN^2\dot y}{y^2}\Bigg) + \Bigg(\frac{\dot\phi^2}{2N} - NV(\phi)\Bigg) \Bigg]y\sqrt{z}dt\end{split}\ee
Now, to study the phase-space structure of action (\ref{A}) for the anisotropic space-time \eqref{aniso}, following Dirac’s algorithm, let us make change of variables, $x = \frac{\dot z}{N}$ and $w = \frac{\dot y}{N}$ in the action (\ref{A2}). Further, treating $(\frac{\dot z}{N} - x) = 0$ and $(\frac{\dot y}{N} - w) = 0$ as constraints, we insert these terms through Lagrange multipliers $\lambda$ and $\tau$ in the point Lagrangian associated with the above action \eqref{A2}, to obtain,

\be\label{L}\begin{split} L = &\Bigg[N\alpha\Bigg(\frac{wx}{yz} + 2\frac{k}{y} - \frac{w^2}{2y^2}\Bigg) - \alpha'\dot\phi\Bigg(\frac{x}{z} + 2\frac{w}{y}\Bigg) + \beta\Bigg(\frac{\dot x^2}{Nz^2} + 4\frac{\dot w\dot x}{Nyz} - \frac{w^2\dot x}{y^2z} + 2\frac{wx\dot x}{yz^2} + 4\frac{\dot w^2}{Ny^2}\\
& + 4\frac{wx\dot w}{y^2z} - 2\frac{x^2\dot w}{yz^2} - \frac{2Nw^3x}{3y^3z}+ \frac{3Nw^2x2}{2y^2z^2} - \frac{2Nwx^3}{3yz^3} - \frac{7Nx^4}{12z^4} + 4\frac{k^2N}{y^2} + 6\frac{kNw^2}{y^3} - \frac{13Nw^4}{12y4}\Bigg)\\
& + \beta'\dot\phi\Bigg(\frac{x^3}{3z^3} -4\frac{kx}{yz}  - 8\frac{kw}{y^2} + \frac{2w^3}{3y^3}\Bigg)  + \gamma'\dot\phi\Bigg(\frac{x^3}{6z^3} - 2\frac{kw}{y^2} + \frac{w^3}{3y^3}\Bigg) + \gamma\Bigg(\frac{\dot x^2}{2Nz^2} + \frac{\dot w\dot x}{Nyz} - \frac{w^2\dot x}{2y^2z}\\
& + \frac{wx\dot x}{2yz^2} + \frac{3\dot w^2}{2Ny^2} + \frac{wx\dot w}{2y^2z} - \frac{x^2\dot w}{2yz^2} + \frac{Nw^3x}{6y^3z} + \frac{5Nw^2x2}{8y^2z^2} - \frac{Nwx^3}{12yz^3} + \frac{7Nx^4}{24z^4} + 2\frac{k^2N}{y^2} + 2\frac{kNw^2}{y^3}\\
& - \frac{5Nw^4}{12y4}\Bigg) + \Bigg(\frac{\dot\phi^2}{2N} - NV\Bigg)\Bigg]y\sqrt{z} + \lambda\Bigg(\frac{\dot z}{N} - x\Bigg) + \tau\Bigg(\frac{\dot y}{N} - w\Bigg).\end{split}\ee
One can clearly observe that the above point Lagrangian \eqref{L} is now cured from the disease of having time derivative of the lapse function $N$. The corresponding momenta are,

\be\label{momenta}\begin{split}& p_x = \frac{\beta}{\sqrt{z}}\Bigg(2\frac{y\dot x}{Nz} + 4\frac{\dot w}{N} + 2\frac{wx}{z} - \frac{w^2}{y}\Bigg) +  \frac{\gamma}{\sqrt{z}}\Bigg(\frac{y\dot x}{Nz} + \frac{\dot w}{N} + \frac{wx}{2z} - \frac{w^2}{2y}\Bigg)\\
& p_w = \frac{\beta}{\sqrt{z}}\Bigg(4\frac{\dot x}{N} + 8\frac{z\dot w}{Ny} + 4\frac{wx}{y} - 2\frac{x^2}{z}\Bigg) +  \frac{\gamma}{\sqrt{z}}\Bigg(\frac{\dot x}{N} + 3\frac{z\dot w}{Ny} + \frac{wx}{2y} - \frac{x^2}{2z}\Bigg)\\
& p_\phi = \Bigg[- \alpha'\Bigg(\frac{x}{z} + 2 \frac{w}{y}\Bigg) + \beta'\Bigg(\frac{x^3}{3z^3} - 4\frac{kx}{yz} + \frac{2w^3}{3y^3} - 8\frac{kw}{y^2}\Bigg) + \gamma'\Bigg(\frac{x^3}{6z^3} - 2\frac{kw}{y^2} + \frac{w^3}{3y^3}\Bigg) + \frac{\dot\phi}{N}\Bigg]y\sqrt{z}\\
& p_z = \frac{\lambda}{N};~~~~p_y = \frac{\tau}{N};~~~~p_\tau = p_N=0=p_\lambda\end{split}\ee
The Hamiltonian constraint therefore reads as,

\be\label{H_c}\begin{split} H_c &= \dot zp_z + \dot xp_x + \dot yp_y + \dot wp_w + \dot \phi p_\phi + \dot Np_N + \dot\lambda p_\lambda + \dot\tau p_\tau - L\\
&=\Bigg[\alpha\left(\frac{w^2}{2y^2} + \frac{wx}{yz} - 2\frac{k}{y}\right)
+ \beta\Bigg(\frac{\dot x^2}{N^2z^2} + 4\frac{\dot w\dot x}{N^2yz} + 4\frac{\dot w^2}{N^2y^2} + \frac{2w^3x}{3y^3z} - \frac{3w^2x^2}{2y^2z^2} + \frac{2wx^3}{3yz^3}\\
&~~~~~ + \frac{7x^4}{12z^4}- 4\frac{k^2}{y^2} - 6\frac{kw^2}{y^3} + \frac{13w^4}{12y^4}\Bigg) + \gamma\Bigg(\frac{\dot x^2}{2N^2z^2} + \frac{\dot w\dot x}{N^2yz} + \frac{3\dot w^2}{2N^2y^2} - \frac{w^3x}{6y^3z} - \frac{5w^2x^2}{8y^2z^2}\\
&~~~~~ + \frac{wx^3}{12yz^3} + \frac{7x^4}{24z^4} - 2\frac{k^2}{y^2} - 2\frac{kw^2}{y^3} + \frac{5w^4}{12y^4}\Bigg) + \left(\frac{\dot\phi^2}{2N^2} + V\right)\Bigg]Ny\sqrt{z} + \lambda x + \tau w.\end{split}\ee
Now, from the expressions of momenta (\ref{momenta}) we find,

\be\label{dot}\begin{split} &\dot x = \frac{Nz}{(3\beta + \gamma)}\Bigg[\frac{\sqrt{z}}{2\gamma}\Bigg\{(8\beta + 3\gamma)\frac{p_x}{y} - (4\beta + \gamma)\frac{p_w}{z}\Bigg\} - 4\frac{\beta^2}{\gamma}\Bigg(\frac{x^2}{z^2} - \frac{w^2}{y^2}\Bigg)\\&
\hspace{1.5 in} - \beta\Bigg(2\frac{x^2}{z^2} + 2\frac{wx}{yz} - \frac{7w^2}{2y^2}\Bigg)- \frac{\gamma}{2}\Bigg(\frac{x^2}{2z^2} + \frac{wx}{yz} - \frac{3w^2}{2y^2}\Bigg)\Bigg];\\
&\dot w = -\frac{Ny}{(3\beta + \gamma)}\Bigg[\frac{\sqrt{z}}{2\gamma}\Bigg\{(4\beta + \gamma)\frac{p_x}{y} - (2\beta + \gamma)\frac{p_w}{z}\Bigg\} - \Bigg(2\frac{\beta^2}{\gamma} + \frac{\gamma}{4}\Bigg)\Bigg(\frac{x^2}{z^2} - \frac{w^2}{y^2}\Bigg) \\&
\hspace{1.2 in}- \frac{\beta}{2}\Bigg(3\frac{x^2}{z^2} - \frac{wx}{yz} - 3\frac{w^2}{y^2}\Bigg)\Bigg];\hspace{0.7 in}
\dot\phi = N\Bigg[\frac{p_\phi}{y\sqrt{z}} + M\Bigg],\end{split}\ee
where, $M=\alpha'\Big(\frac{x}{z} + 2 \frac{w}{y}\Big) - \beta'\Big(\frac{x^3}{3z^3} - 4\frac{kx}{yz} + \frac{2w^3}{3y^3} - 8\frac{kw}{y^2}\Big) - \gamma'\Big(\frac{x^3}{6z^3} - 2\frac{kw}{y^2} + \frac{w^3}{3y^3}\Big).$ Using these expressions (\ref{dot}), it is now possible to express the Hamiltonian (\ref{H_c}) as,

\be\label{H_c1}\begin{split} &H_c = N\Bigg[\frac{1}{2(3\beta + \gamma)}\Bigg\{(8\beta + 3\gamma)\frac{z^{\frac{3}{2}}p_x^2}{2y\gamma} + (2\beta + \gamma)\frac{yp_w^2}{2\sqrt{z}\gamma} - \frac{4\beta + \gamma}{\gamma}\sqrt{z}p_xp_w - zp_x\Bigg((4\beta + \gamma)^2\frac{x^2}{2z^2\gamma}\\
& + (4\beta + \gamma)\frac{wx}{yz} - (2\beta + \gamma)(8\beta + 3\gamma)\frac{w^2}{2y^2\gamma}\Bigg) + yp_w\Bigg((2\beta + \gamma)(4\beta + \gamma)\frac{1}{2\gamma}\Big(\frac{x^2}{z^2} - \frac{w^2}{y^2}\Big) - \frac{wx\beta}{yz}\Bigg)\Bigg\}\\
& + \frac{p_\phi^2}{2y\sqrt{z}} + M p_\phi + \Bigg\{\frac{M^2}{2} + \frac{1}{3\beta + \gamma}\Bigg(2\frac{\beta^3}{\gamma}\Big(\frac{x^2}{z^2} - \frac{w^2}{y^2}\Big)^2 + \beta^2\Big(\frac{15x^4}{4z^4} + \frac{x^3w}{yz^3} - \frac{11x^2w^2}{2y^2z^2} + 6\frac{w^4}{y^4}\Big) +\\
&\beta\gamma\Big(\frac{25x^4}{12z^4} + \frac{2x^3w}{3yz^3} - \frac{7x^2w^2}{2y^2z^2} - \frac{4xw^3}{3y^3z} + \frac{43w^4}{12y^4}\Big) + \frac{\gamma^2}{4}\Big(\frac{17x^4}{12z^4} + \frac{x^3w}{3yz^3} - \frac{5x^2w^2}{2y^2z^2} - \frac{5xw^3}{3y^3z} + \frac{29w^4}{12y^4}\Big)\Bigg)\\
& + \alpha\Bigg(\frac{w^2}{2y^2} + \frac{wx}{yz} - 2\frac{k}{y}\Bigg) - 2k\beta\Bigg(3\frac{w^2}{y^3} + 2\frac{k}{y^2}\Bigg) - 2k\gamma\Bigg(\frac{w^2}{y^3} + \frac{k}{y^2}\Bigg) + V\Bigg\}y\sqrt{z}\Bigg] + \lambda x + \tau w.\end{split}\ee
In view of the definition of momenta \eqref{momenta}, we therefore require five primary constraints involving Lagrange multiplier or its conjugate viz, $\phi_1 = Np_z - \lambda \approx 0,~ \phi_2 = Np_y - \tau \approx 0,~ \phi_3 = p_\lambda \approx 0,~ \phi_4 = p_\tau \approx 0, ~\mathrm{and,} ~\phi_5 = p_N \approx 0$, which are all second class constraints, since, $\{\phi_i, \phi_j\} \ne 0$. Note that, since the lapse function $N$ is non-dynamical, so the associated constraint vanishes strongly, and therefore it may be safely ignored. The first four second class constraints may now be harmlessly substituted and the modified primary Hamiltonian takes the form,

\be H_{p1} = H_c + u_1(Np_z - \lambda) + u_2p_\lambda + u_3(Np_y - \tau) + u_4p_\tau\ee
In the above, $u_1$, $u_2$, $u_3$ and $u_4$ are Lagrange multipliers, and the Poisson brackets $\{x, p_x\} = \{z, p_z\} = \{\lambda, p_\lambda\} = \{w, p_w\} = \{y, p_y\} = \{\tau, p_\tau\} = 1$, hold. The requirement that the constraints must remain preserved in time is exhibited in the Poisson brackets $\{\phi_i, H_{p1}\}$ viz,

\be\begin{split} &\dot\phi_1 = \{\phi_1,H_{p1}\} = -N \frac{\partial H_{p1}}{\partial z} - u_2 + \Sigma_{i=1}^2\phi_i\{\phi_1,u_i\},\\
&\dot\phi_2 = \{\phi_2,H_{p1}\} = x - u_1 + \Sigma_{i=1}^2\phi_i\{\phi_2,u_i\},\\
&\dot\phi_3 = \{\phi_3,H_{p1}\} = -N \frac{\partial H_{p1}}{\partial y} - u_4 + \Sigma_{i=1}^2\phi_i\{\phi_3,u_i\},\\
&\dot\phi_4 = \{\phi_4,H_{p1}\} = w - u_3 + \Sigma_{i=1}^2\phi_i\{\phi_4,u_i\}.\end{split}\ee
Now, constraints must also vanish weakly in the sense of Dirac. As a result, $\{\phi_1, H_{p1}\} = \dot\phi_1 \approx 0$, leads to, $u_2 = - N \frac{\partial H_{p1}}{\partial z}$ , $\{\phi_2, H_{p1}\} = \dot\phi_2 \approx 0$, leads to, $u_1 = x$, $\{\phi_3, H_{p1}\} = \dot\phi_3 \approx 0$, leads to, $u_4 = - N \frac{\partial H_{p1}}{\partial y}$ , and $\{\phi_4, H_{p1}\} = \dot\phi_4 \approx 0$, leads to, $u_3 = w$. On thus imposing these conditions, $H_{p1}$ is then modified by the primary Hamiltonian $H_{p2}$, which now reads as,

\be\begin{split} &H_{p2} =  N\Bigg[ xp_z + wp_y + \frac{1}{2(3\beta + \gamma)}\Bigg\{(8\beta + 3\gamma)\frac{z^{\frac{3}{2}}p_x^2}{2y\gamma} + (2\beta + \gamma)\frac{yp_w^2}{2\sqrt{z}\gamma} - \frac{4\beta + \gamma}{\gamma}\sqrt{z}p_xp_w\\
&~~ - zp_x\Bigg((4\beta + \gamma)^2\frac{x^2}{2z^2\gamma} + (4\beta + \gamma)\frac{wx}{yz} - (2\beta + \gamma)(8\beta + 3\gamma)\frac{w^2}{2y^2\gamma}\Bigg) + yp_w\Bigg( - \frac{wx\beta}{yz}\\
&~~ + (2\beta + \gamma)(4\beta + \gamma)\frac{1}{2\gamma}\Big(\frac{x^2}{z^2} - \frac{w^2}{y^2}\Big)\Bigg)\Bigg\} + \frac{p_\phi^2}{2y\sqrt{z}} + M p_\phi + \Bigg\{\frac{M^2}{2} + \frac{1}{3\beta + \gamma}\Bigg(2\frac{\beta^3}{\gamma}\Big(\frac{x^2}{z^2} - \frac{w^2}{y^2}\Big)^2\\
&~~ + \beta^2\Big(\frac{15x^4}{4z^4} + \frac{x^3w}{yz^3} - \frac{11x^2w^2}{2y^2z^2} + 6\frac{w^4}{y^4}\Big) + \beta\gamma\Big(\frac{25x^4}{12z^4} + \frac{2x^3w}{3yz^3} - \frac{7x^2w^2}{2y^2z^2} - \frac{4xw^3}{3y^3z} + \frac{43w^4}{12y^4}\Big)\\
& + \frac{\gamma^2}{4}\Big(\frac{17x^4}{12z^4} + \frac{x^3w}{3yz^3} - \frac{5x^2w^2}{2y^2z^2} - \frac{5xw^3}{3y^3z}  + \frac{29w^4}{12y^4}\Big)\Bigg) + \alpha\Bigg(\frac{w^2}{2y^2} + \frac{wx}{yz} - 2\frac{k}{y}\Bigg) - 2k\beta\Bigg(3\frac{w^2}{y^3} + 2\frac{k}{y^2}\Bigg)\\
&~~ - 2k\gamma\Bigg(\frac{w^2}{y^3} + \frac{k}{y^2}\Bigg) + V\Bigg\}y\sqrt{z} - \frac{\partial H_{p1}}{\partial z}p_\lambda - \frac{\partial H_{p1}}{\partial y}p_\tau\Bigg]\end{split}\ee
We repeat that constraints must vanish weakly in the sense of Dirac, and therefore in view of the Poisson brackets $\{\phi_1, H_{p2}\} = \dot\phi_1 \approx 0$ and $\{\phi_3, H_{p2}\} = \dot\phi_3 \approx 0$, one obtains $p_\lambda = 0$ and $p_\tau = 0$, respectively. Thus the Hamiltonian finally takes the form $H_D = N\mathcal{H_D}$, where,

\be\label{HD}\begin{split} \mathcal{H_D} =& xp_z + wp_y + \frac{1}{2(3\beta + \gamma)}\Bigg\{(8\beta + 3\gamma)\frac{z^{\frac{3}{2}}p_x^2}{2y\gamma} + (2\beta + \gamma)\frac{yp_w^2}{2\sqrt{z}\gamma} - \frac{4\beta + \gamma}{\gamma}\sqrt{z}p_xp_w\\
&~~ - zp_x\Bigg((4\beta + \gamma)^2\frac{x^2}{2z^2\gamma} + (4\beta + \gamma)\frac{wx}{yz} - (2\beta + \gamma)(8\beta + 3\gamma)\frac{w^2}{2y^2\gamma}\Bigg)\\
&~~ + yp_w\Bigg((2\beta + \gamma)(4\beta + \gamma)\frac{1}{2\gamma}\Big(\frac{x^2}{z^2} - \frac{w^2}{y^2}\Big) - \frac{wx\beta}{yz}\Bigg)\Bigg\} + \frac{p_\phi^2}{2y\sqrt{z}} + M p_\phi\\
&~~ + \Bigg\{\frac{M^2}{2} + \frac{1}{3\beta + \gamma}\Bigg(2\frac{\beta^3}{\gamma}\Big(\frac{x^2}{z^2} - \frac{w^2}{y^2}\Big)^2 + \beta^2\Big(\frac{15x^4}{4z^4} + \frac{x^3w}{yz^3} - \frac{11x^2w^2}{2y^2z^2} + 6\frac{w^4}{y^4}\Big)\\
& + \beta\gamma\Big(\frac{25x^4}{12z^4} + \frac{2x^3w}{3yz^3} - \frac{7x^2w^2}{2y^2z^2} - \frac{4xw^3}{3y^3z} + \frac{43w^4}{12y^4}\Big) + \frac{\gamma^2}{4}\Big(\frac{17x^4}{12z^4} + \frac{x^3w}{3yz^3} - \frac{5x^2w^2}{2y^2z^2} - \frac{5xw^3}{3y^3z}\\
& + \frac{29w^4}{12y^4}\Big)\Bigg) + \alpha\Bigg(\frac{w^2}{2y^2} + \frac{wx}{yz} - 2\frac{k}{y}\Bigg) - 2k\beta\Bigg(3\frac{w^2}{y^3} + 2\frac{k}{y^2}\Bigg) - 2k\gamma\Bigg(\frac{w^2}{y^3} + \frac{k}{y^2}\Bigg) + V\Bigg\}y\sqrt{z},\end{split}\ee
and in the process diffeomorphic invariance is clearly established. Now, since $\dot z = Nx$ and $\dot y = Nw$, we have $\dot z p_z + \dot y p_y = N(xp_z + wp_y)$. So, using the expressions of $p_x, p_w, p_\phi$ from equation (\ref{momenta}) and $\mathcal{H_D}$ from equation (\ref{HD}), it is now straightforward to write the action (\ref{A2}) in the following canonical (ADM) form as,

\be A = \int(\dot zp_z + \dot xp_x + \dot yp_y + \dot wp_w + \dot\phi p_\phi - N\mathcal{H_D})dt,\ee
which amounts in writing,

\be \label{ADM} A = \int(\dot h_{ij}p^{ij} + \dot K_{ij}\pi^{ij} + \dot\phi p_\phi - N\mathcal{H_D})dt.\ee

\subsection{Modified Horowitz formalism:}

In this subsection, we seek the phase-space structure of action (\ref{A}), once again following Modified Horowitz' formalism. As mentioned, we first need to control the divergent terms, upon integrating the action (\ref{A1}) by parts, to end up with the action (\ref{A2}), which is our starting point as before. Now substituting the auxiliary variables, $Q_1$, and $Q_2$,

\be\begin{split} &Q_1 = N {\partial A\over \partial \ddot z}\\& = \frac{y}{N^2\sqrt{z}}\Bigg[\beta\Bigg(2\frac{\ddot z}{z} + 4\frac{\ddot y}{y} + 2\frac{\dot z\dot y}{zy} - \frac{\dot y^2}{y^2} - 2\frac{\dot N\dot z}{Nz} - 4\frac{\dot N\dot y}{Ny}\Bigg) + \gamma\Bigg(\frac{\ddot z}{z} + \frac{\ddot y}{y} + \frac{\dot z\dot y}{2zy} - \frac{\dot y^2}{2y^2} - \frac{\dot N\dot z}{Nz} - \frac{\dot N\dot y}{Ny}\Bigg)\Bigg]\\
&Q_2 = N {\partial A\over \partial \ddot y}\\& = \frac{\sqrt{z}}{N^2}\Bigg[2\beta\Bigg(2\frac{\ddot z}{z} + 4\frac{\ddot y}{y} + 2\frac{\dot z\dot y}{zy} - \frac{\dot z^2}{z^2} - 2\frac{\dot N\dot z}{Nz} - 4\frac{\dot N\dot y}{Ny}\Bigg) + \gamma\Bigg(\frac{\ddot z}{z} + 3\frac{\ddot y}{y} + \frac{\dot z\dot y}{2zy} - \frac{\dot z^2}{2z^2} - \frac{\dot N\dot z}{Nz} - 3\frac{\dot N\dot y}{Ny}\Bigg)\Bigg]\end{split}\ee
judiciously into the action (\ref{A2}) one obtains,

\be\label{AM1}\begin{split} A &= \int\Bigg[\frac{Q_1\ddot z}{N} + \frac{Q_2\ddot y}{N} - \frac{Q_1\dot N\dot z}{N^2} - \frac{Q_2\dot N\dot y}{N^2} - \frac{1}{2(3\beta + \gamma)}\Bigg\{(8\beta + 3\gamma)\frac{Nz^{\frac{3}{2}}Q_1^2}{2y\gamma} + (2\beta + \gamma)\frac{NyQ_2^2}{2\sqrt{z}\gamma}\\
& - \frac{4\beta + \gamma}{\gamma}N\sqrt{z}Q_1Q_2 - \frac{zQ_1}{N}\Bigg((4\beta + \gamma)^2\frac{\dot z^2}{2z^2\gamma} + (4\beta + \gamma)\frac{\dot z\dot y}{zy} - (2\beta + \gamma)(8\beta + 3\gamma)\frac{\dot y^2}{2y^2\gamma}\Bigg)\\
& + \frac{yQ_2}{N}\Bigg((2\beta + \gamma)(4\beta + \gamma)\frac{1}{2\gamma}\Big(\frac{\dot z^2}{z^2} - \frac{\dot y^2}{y^2}\Big) - \frac{\dot z\dot y\beta}{yz}\Bigg)\Bigg\} + \Bigg\{-\alpha'\dot\phi\Bigg(\frac{\dot z}{z} + 2\frac{\dot y}{y}\Bigg) + \frac{\beta'\dot\phi}{N^2}\Bigg(\frac{\dot z^3}{3z^3}\\
& + \frac{2\dot y^3}{3y^3} - 4\frac{kN^2\dot z}{zy} - 8\frac{kN^2\dot y}{y^2}\Bigg) + \frac{\gamma'\dot\phi}{N^2}\Bigg(\frac{\dot z^3}{6z^3} + \frac{\dot y^3}{3y^3} - 2\frac{kN^2\dot y}{y^2}\Bigg) + \frac{1}{N^2(3\beta + \gamma)}\Bigg(2\frac{\beta^3}{\gamma}\Big(\frac{\dot z^2}{z^2} - \frac{\dot y^2}{y^2}\Big)^2\\
& + \beta^2\Big(\frac{15\dot z^4}{4z^4} - \frac{11\dot y^2\dot z^2}{2y^2z^2} + \frac{\dot y\dot z^3}{yz^3} + \frac{3\dot y^4}{2y^4}\Big) + \beta\gamma\Big(\frac{25\dot z^4}{12z^4} - \frac{4\dot y^3\dot z}{3y^3z} - \frac{7\dot y^2\dot z^2}{2y^2z^2} + \frac{2\dot y\dot z^3}{3yz^3} + \frac{43\dot y^4}{12y^4}\Big)\\
& + \frac{\gamma^2}{4}\Big(\frac{17\dot z^4}{12z^4} - \frac{5\dot y^3\dot z}{3y^3z} - \frac{5\dot y^2\dot z^2}{2y^2z^2} + \frac{\dot y\dot z^3}{3yz^3} + \frac{29\dot y^4}{12y^4}\Big)\Bigg) - \alpha\Big(\frac{\dot y^2}{2y^2} + \frac{\dot y\dot z}{yz} - 2\frac{kN^2}{y}\Big) + 2k\beta\Big(2\frac{kN^2}{y^2}\\
& + 3\frac{\dot y^2}{y^3}\Big) + 2k\gamma\Big(\frac{k^2N^2}{y^2} + \frac{\dot y^2}{y^3}\Big) + \frac{\dot\phi^2}{2} - N^2V\Bigg\}\frac{y\sqrt{z}}{N}\Bigg]dt.\end{split}\ee
The rest of the total derivative terms, viz., $\big(\frac{Q_1\dot z}{N} + \frac{Q_2\dot y}{N}\big)$ are taken care of, under integration by parts yet again, and finally the action \eqref{AM1} is expressed as,

\be\label{AM2}\begin{split} &A = \int\Bigg[-\frac{\dot Q_1\dot z}{N} - \frac{\dot Q_2\dot y}{N} - \frac{1}{2(3\beta + \gamma)}\Bigg\{(8\beta + 3\gamma)\frac{Nz^{\frac{3}{2}}Q_1^2}{2y\gamma} + (2\beta + \gamma)\frac{NyQ_2^2}{2\sqrt{z}\gamma} - \frac{4\beta + \gamma}{\gamma}N\sqrt{z}Q_1Q_2\\
& - \frac{zQ_1}{N}\Bigg((4\beta + \gamma)^2\frac{\dot z^2}{2z^2\gamma} + (4\beta + \gamma)\frac{\dot z\dot y}{zy} - (2\beta + \gamma)(8\beta + 3\gamma)\frac{\dot y^2}{2y^2\gamma}\Bigg) + \frac{yQ_2}{N}\Bigg( - \frac{\dot z\dot y\beta}{yz} ~+\\
& (2\beta + \gamma)(4\beta + \gamma)\frac{1}{2\gamma}\Big(\frac{\dot z^2}{z^2} - \frac{\dot y^2}{y^2}\Big)\Bigg)\Bigg\} + \Bigg\{ \frac{\beta'\dot\phi}{N^2}\Bigg(\frac{\dot z^3}{3z^3} + \frac{2\dot y^3}{3y^3} - 4\frac{kN^2}{y}\Big(\frac{\dot z}{z} + 2\frac{\dot y}{y}\Big)\Bigg) -\alpha'\dot\phi\Bigg(\frac{\dot z}{z} + 2\frac{\dot y}{y}\Bigg) + \\
& \frac{\gamma'\dot\phi}{N^2}\Bigg(\frac{\dot z^3}{6z^3} + \frac{\dot y^3}{3y^3} - \frac{2kN^2\dot y}{y^2}\Bigg) + \frac{1}{N^2(3\beta + \gamma)}\Bigg(\frac{2\beta^3}{\gamma}\Big(\frac{\dot z^2}{z^2} - \frac{\dot y^2}{y^2}\Big)^2 + \beta^2\Big(\frac{15\dot z^4}{4z^4} - \frac{11\dot y^2\dot z^2}{2y^2z^2} + \frac{\dot y\dot z^3}{yz^3} + \frac{3\dot y^4}{2y^4}\Big)\\
& + \beta\gamma\Big(\frac{25\dot z^4}{12z^4} - \frac{4\dot y^3\dot z}{3y^3z} - \frac{7\dot y^2\dot z^2}{2y^2z^2} + \frac{2\dot y\dot z^3}{3yz^3} + \frac{43\dot y^4}{12y^4}\Big) + \frac{\gamma^2}{4}\Big(\frac{17\dot z^4}{12z^4} - \frac{5\dot y^3\dot z}{3y^3z} - \frac{5\dot y^2\dot z^2}{2y^2z^2} + \frac{\dot y\dot z^3}{3yz^3} + \frac{29\dot y^4}{12y^4}\Big)\Bigg)\\
& - \alpha\Big(\frac{\dot y^2}{2y^2} + \frac{\dot y\dot z}{yz} - 2\frac{kN^2}{y}\Big) + 2k\beta\Big(2\frac{kN^2}{y^2} + 3\frac{\dot y^2}{y^3}\Big) + 2k\gamma\Big(\frac{k^2N^2}{y^2} + \frac{\dot y^2}{y^3}\Big) + \frac{\dot\phi^2}{2} - N^2V\Bigg\}\frac{y\sqrt{z}}{N}\Bigg]dt\end{split}\ee
The following canonical momenta,

\be\label{momenta MHF}\begin{split}
& p_z =  - \frac{\dot Q_1}{N} + \frac{1}{2N(3\beta + \gamma)}\Bigg\{(4\beta + \gamma) Q_1\Bigg((4\beta + \gamma)\frac{\dot z}{z\gamma} + \frac{\dot y}{y}\Bigg) + \frac{yQ_2}{z}\Bigg(\beta\frac{\dot y}{y} - (2\beta + \gamma)(4\beta + \gamma)\frac{\dot z}{z\gamma}\Bigg)\Bigg\}\\
&~ + \Bigg\{\frac{\beta'\dot\phi}{N^2z}\Bigg(\frac{\dot z^2}{z^2} - 4\frac{kN^2}{y}\Bigg) + \frac{\gamma'\dot\phi\dot z^2}{2N^2z^3} - \frac{\alpha'\dot\phi}{z} - \alpha\frac{\dot y}{yz} + \frac{1}{N^2z(3\beta + \gamma)}\Bigg(8\frac{\beta^3\dot z}{\gamma z}\Big(\frac{\dot z^2}{z^2} - \frac{\dot y^2}{y^2}\Big)^2 + \beta^2\Big(15\frac{\dot z^3}{z^3}\\
& - 11\frac{\dot y^2\dot z}{y^2z} + 3\frac{\dot y\dot z^2}{yz^2}\Big) + \beta\gamma\Big(\frac{25\dot z^3}{3z^3} - \frac{4\dot y^3}{3y^3} - 7\frac{\dot y^2\dot z}{y^2z} + 2\frac{\dot y\dot z^2}{yz^2}\Big) + \frac{\gamma^2}{4}\Big(\frac{17\dot z^3}{3z^3} - \frac{5\dot y^3}{3y^3} - 5\frac{\dot y^2\dot z}{y^2z} + \frac{\dot y\dot z^2}{yz^2}\Big)\Bigg)\Bigg\}\frac{y\sqrt{z}}{N}\\
& p_y = \frac{1}{2N(3\beta + \gamma)}\Bigg\{\frac{zQ_1}{y}\Big((4\beta + \gamma)\frac{\dot z}{z} - (2\beta + \gamma)(8\beta + 3\gamma)\frac{\dot y}{y\gamma}\Big) + Q_2\Big((2\beta + \gamma)(4\beta + \gamma)\frac{\dot y}{\gamma y} + \frac{\dot z\beta}{z}\Big)\Bigg\}\\
&\hspace{1.2in} - \frac{\dot Q_2}{N} + \Bigg\{ - 2\alpha'\dot\phi + 2\frac{\beta'\dot\phi}{N^2}\Bigg(\frac{\dot y^2}{y^2} - 4\frac{kN^2}{y^2}\Bigg) + \frac{\gamma'\dot\phi}{N^2}\Bigg(\frac{\dot y^2}{y^2} - 2\frac{kN^2}{y^2}\Bigg) - \alpha\Bigg(\frac{\dot y}{y} + \frac{\dot z}{z}\Bigg)\\
&\hspace{1.2in} + 4k(3\beta + \gamma)\frac{\dot y}{y^2} + \frac{1}{N^2(3\beta + \gamma)}\Bigg(8\frac{\beta^3}{\gamma}\frac{\dot y}{y}\Big(\frac{\dot y^2}{y^2} - \frac{\dot z^2}{z^2}\Big) + \beta^2\Big(\frac{\dot z^3}{z^3} - 11\frac{\dot y\dot z^2}{yz^2}  + 24\frac{\dot y^3}{y^3}\Big)\\
&\hspace{1.2in} + \beta\gamma\Big(\frac{2\dot z^3}{3z^3} - 4\frac{\dot y^2\dot z}{y^2z} - 7\frac{\dot y\dot z^2}{yz^2} + \frac{43\dot y^3}{3y^3}\Big) + \frac{\gamma^2}{4}\Big(\frac{\dot z^3}{3z^3} - 5\frac{\dot y^2\dot z}{y^2z} - 5\frac{\dot y\dot z^2}{yz^2} + \frac{29\dot y^3}{3y^3}\Big)\Bigg)\Bigg\}\frac{\sqrt{z}}{N}\\
& p_\phi = \Bigg\{\dot\phi-\alpha'\Bigg(\frac{\dot z}{z} + 2\frac{\dot y}{y}\Bigg) + \frac{\beta'}{N^2}\Bigg(\frac{\dot z^3}{3z^3} + \frac{2\dot y^3}{3y^3} - 4\frac{kN^2\dot z}{zy} - 8\frac{kN^2\dot y}{y^2}\Bigg)\\
&\hspace{0.25in} + \frac{\gamma'}{N^2}\Bigg(\frac{\dot z^3}{6z^3} + \frac{\dot y^3}{3y^3}  - 2\frac{kN^2\dot y}{y^2}\Bigg)\Bigg\}\frac{y\sqrt{z}}{N};\hspace{0.25in}p_{Q_1} = - \frac{\dot z}{N};\hspace{0.25in}p_{Q_2} = - \frac{\dot y}{N};\hspace{0.25in}p_N = 0,\end{split}\ee
clearly signal that the action is degenerate. The constraint Hamiltonian may now be expressed as,

\be\label{HM}\begin{split} &H_c = \dot zp_z + \dot Q_1p_{Q_1} + \dot yp_y + \dot Q_2p_{Q_2} + \dot \phi p_\phi + \dot Np_N  - L\\
&~= -\frac{\dot Q_1\dot z}{N} - \frac{\dot Q_2\dot y}{N} + \frac{1}{2(3\beta + \gamma)}\Bigg\{(8\beta + 3\gamma)\frac{Nz^{\frac{3}{2}}Q_1^2}{2y\gamma} + (2\beta + \gamma)\frac{NyQ_2^2}{2\sqrt{z}\gamma} - \frac{4\beta + \gamma}{\gamma}N\sqrt{z}Q_1Q_2\\
&~ + \frac{zQ_1}{N}\Bigg((4\beta + \gamma)^2\frac{\dot z^2}{2z^2\gamma} + (4\beta + \gamma)\frac{\dot z\dot y}{zy} - (2\beta + \gamma)(8\beta + 3\gamma)\frac{\dot y^2}{2y^2\gamma}\Bigg) - \frac{yQ_2}{N}\Bigg(- \frac{\dot z\dot y\beta}{yz} \\
& + (2\beta + \gamma)(4\beta + \gamma)\frac{1}{2\gamma}\Big(\frac{\dot z^2}{z^2} - \frac{\dot y^2}{y^2}\Big) \Bigg)\Bigg\} + \Bigg\{ \frac{\beta'\dot\phi}{N^2}\Bigg(\frac{\dot z^3}{z^3} + 2\frac{\dot y^3}{y^3} - 4\frac{kN^2}{y}\Big(\frac{\dot z}{z} + 2\frac{\dot y}{y}\Big)\Bigg) - \alpha'\dot\phi\Bigg(\frac{\dot z}{z} + 2\frac{\dot y}{y}\Bigg) \\
&~ + \frac{\gamma'\dot\phi}{N^2}\Bigg(\frac{\dot z^3}{2z^3} + \frac{\dot y^3}{y^3}  - 2\frac{kN^2\dot y}{y^2}\Bigg) - \frac{1}{N^2(3\beta + \gamma)}\Bigg(6\frac{\beta^3}{\gamma}\Big(\frac{\dot z^2}{z^2} - \frac{\dot y^2}{y^2}\Big)^2 + 3\beta^2\Big(\frac{15\dot z^4}{4z^4} - \frac{11\dot y^2\dot z^2}{2y^2z^2} + \frac{\dot y\dot z^3}{yz^3}\\
&~ + 6\frac{\dot y^4}{y^4}\Big) + \beta\gamma\Big(\frac{25\dot z^4}{4z^4} - 4\frac{\dot y^3\dot z}{y^3z} - \frac{21\dot y^2\dot z^2}{2y^2z^2} + \frac{2\dot y\dot z^3}{yz^3} + \frac{43\dot y^4}{4y^4}\Big) + \frac{\gamma^2}{4}\Big(\frac{17\dot z^4}{4z^4} - 5\frac{\dot y^3\dot z}{y^3z} - \frac{15\dot y^2\dot z^2}{2y^2z^2} + \frac{\dot y\dot z^3}{yz^3}\\
&~ + \frac{29\dot y^4}{4y^4}\Big)\Bigg) - \alpha\Big(\frac{\dot y^2}{2y^2} + \frac{\dot y\dot z}{yz} + 2\frac{kN^2}{y}\Big) + 2k(3\beta + \gamma)\frac{\dot y^2}{y^3} + \frac{\dot\phi^2}{2} + N^2V\Bigg\}\frac{y\sqrt{z}}{N}.\end{split}\ee
Now, finding the expressions $p_{Q_1}p_z$ and $p_{Q_2}p_y$ in view of the definitions of momenta (\ref{momenta MHF}) and putting the same into the constraint Hamiltonian (\ref{HM}), it is possible to bypass Dirac's constraint analysis since, we can express the constraint Hamiltonian (\ref{HM}) in terms of the phase-space variables as,

\be\label{HM1}\begin{split} &H_M = N\mathcal{H_M} = N\Bigg[- p_{Q_1}p_z - p_{Q_2}p_y + \frac{1}{2(3\beta + \gamma)}\Bigg\{(8\beta + 3\gamma)\frac{z^{\frac{3}{2}}Q_1^2}{2y\gamma} + (2\beta + \gamma)\frac{yQ_2^2}{2\sqrt{z}\gamma}\\
&~~ - \frac{4\beta + \gamma}{\gamma}\sqrt{z}Q_1Q_2 - \frac{zQ_1}{2}\Bigg((4\beta + \gamma)^2\frac{p_{Q_1}^2}{2z^2\gamma} + (4\beta + \gamma)\frac{p_{Q_1}p_{Q_2}}{zy} - (2\beta + \gamma)(8\beta + 3\gamma)\frac{p_{Q_2}^2}{2y^2\gamma}\Bigg)\\
&~~ + \frac{yQ_2}{2}\Bigg((2\beta + \gamma)(4\beta + \gamma)\frac{1}{2\gamma}\Bigg(\frac{p_{Q_1}^2}{z^2} - \frac{p_{Q_2}^2}{y^2}\Bigg) - \beta\frac{p_{Q_1}p_{Q_2}}{yz}\Bigg)\Bigg\} + \frac{p_\phi^2}{2y\sqrt{z}} - Up_\phi + \Bigg\{\frac{U^2}{2}\\
&~~ + \frac{1}{3\beta + \gamma}\Bigg(2\frac{\beta^3}{\gamma}\Big(\frac{p_{Q_1}^2}{z^2} - \frac{p_{Q_2}^2}{y^2}\Big)^2 + \beta^2\Big(\frac{15p_{Q_1}^4}{4z^4} + \frac{p_{Q_1}^3p_{Q_2}}{yz^3} - \frac{11p_{Q_1}^2p_{Q_2}^2}{2y^2z^2} + 6\frac{p_{Q_2}^4}{y^4}\Big) + \beta\gamma\Big(\frac{25p_{Q_1}^4}{12z^4}\\
&~~ + \frac{2p_{Q_1}^3p_{Q_2}}{3yz^3} - \frac{7p_{Q_1}^2p_{Q_2}^2}{2y^2z^2} - \frac{4p_{Q_1}p_{Q_2}^3}{3y^3z} + \frac{43p_{Q_2}^4}{12y^4}\Big) + \frac{\gamma^2}{4}\Big(\frac{17p_{Q_1}^4}{12z^4} + \frac{p_{Q_1}^3p_{Q_2}}{3yz^3} - \frac{5p_{Q_1}^2p_{Q_2}^2}{2y^2z^2} - \frac{5p_{Q_1}p_{Q_2}^3}{3y^3z}\\
&~~ + \frac{29p_{Q_2}^4}{12y^4}\Big)\Bigg) + \alpha\Bigg(\frac{p_{Q_2}^2}{2y^2} + \frac{p_{Q_1}p_{Q_2}}{yz} - 2\frac{k}{y}\Bigg) - 2k\beta\Bigg(3\frac{p_{Q_2}^2}{y^3} + 2\frac{k}{y^2}\Bigg) - 2k\gamma\Bigg(\frac{p_{Q_2}^2}{y^3} + \frac{k}{y^2}\Bigg) + V\Bigg\}y\sqrt{z}\Bigg]\end{split}\ee
where, $U = \alpha'\Big(\frac{p_{Q_1}}{z} + 2\frac{p_{Q_2}}{y}\Big) - \beta'\Big(\frac{p_{Q_1}^3}{3z^3} + \frac{2p_{Q_2}^3}{3y^3} - 4\frac{kp_{Q_1}}{yz} - 8\frac{kp_{Q_2}}{y^2}\Big) - \gamma'\Big(\frac{p_{Q_1}^3}{6z^3} + \frac{p_{Q_2}^3}{3y^3}  - 2\frac{kp_{Q_2}}{y^2}\Big)$, and in the process, diffeomorphic invariance is established. However, the appearance of momenta ${P_Q}_1$, and ${P_Q}_2$ with fourth degree desist the Hamiltonian from casting a viable quantum dynamics. Thus, we now express the Hamiltonian in terms of the basic variables. This is performed by replacing $Q_1$ by $p_x$, $Q_2$ by $p_w$, $p_{Q_1}$ by $-x$ and $p_{Q_2}$ by $-w$. These indeed are canonical transformations, since $p_{Q_1} = - \frac{\dot z}{N} = -x$ and $Q_1 = N\frac{\partial A}{\partial\ddot z} = N\frac{\partial A}{\partial\dot x}\frac{\partial \dot x}{\partial\ddot z} = N\times p_x\times\frac{1}{N} = p_x$, thus, $\{x,p_x\} = \frac{\partial x}{\partial Q_1}\frac{\partial p_x}{\partial p_{Q_1}} - \frac{\partial x}{\partial p_{Q_1}}\frac{\partial p_x}{\partial Q_1} = 0 - (-1)\times 1=1$. Similarly, $p_{Q_2} = - \frac{\dot y}{N} = -w$ and $Q_2 = N\frac{\partial A}{\partial\ddot y} = N\frac{\partial A}{\partial\dot w}\frac{\partial \dot w}{\partial\ddot z} = p_w$, and thus, $\{w,p_w\} =1$. The Hamiltonian therefore finally reads as,

\be\label{HM2}\begin{split} &\mathcal{H_M} =  xp_z + wp_y + \frac{1}{2(3\beta + \gamma)}\Bigg\{(8\beta + 3\gamma)\frac{z^{\frac{3}{2}}p_x^2}{2y\gamma} + (2\beta + \gamma)\frac{yp_w^2}{2\sqrt{z}\gamma} - \frac{4\beta + \gamma}{\gamma}\sqrt{z}p_xp_w - \frac{zp_x}{2}\\
&~~\Bigg((4\beta + \gamma)^2\frac{x^2}{2z^2\gamma} + (4\beta + \gamma)\frac{xw}{zy} - (2\beta + \gamma)(8\beta + 3\gamma)\frac{w^2}{2y^2\gamma}\Bigg) + \frac{yp_w}{2}\Bigg( - \beta\frac{xw}{yz} + (2\beta + \gamma)(4\beta + \gamma)\\
&~~\frac{1}{2\gamma}\Bigg(\frac{x^2}{z^2} - \frac{w^2}{y^2}\Bigg)\Bigg)\Bigg\} + \frac{p_\phi^2}{2y\sqrt{z}} + Mp_\phi + \Bigg\{\frac{M^2}{2} + \frac{1}{3\beta + \gamma}\Bigg(2\frac{\beta^3}{\gamma}\Big(\frac{x^2}{z^2} - \frac{w^2}{y^2}\Big)^2 + \beta^2\Big(\frac{15x^4}{4z^4} + \frac{x^3w}{yz^3}\\
& - \frac{11x^2w^2}{2y^2z^2} + 6\frac{w^4}{y^4}\Big) + \beta\gamma\Big(\frac{25x^4}{12z^4} + \frac{2x^3w}{3yz^3} - \frac{7x^2w^2}{2y^2z^2} - \frac{4xw^3}{3y^3z} + \frac{43w^4}{12y^4}\Big) + \frac{\gamma^2}{4}\Big(\frac{17x^4}{12z^4} + \frac{x^3w}{3yz^3} - \frac{5x^2w^2}{2y^2z^2}\\
& - \frac{5xw^3}{3y^3z}  + \frac{29w^4}{12y^4}\Big)\Bigg) + \alpha\Big(\frac{w^2}{2y^2} + \frac{wx}{yz} - \frac{2k}{y}\Big) - 2k\beta\Big(\frac{3w^2}{y^3} + \frac{2k}{y^2}\Big) - 2k\gamma\Big(\frac{w^2}{y^3} + \frac{k}{y^2}\Big) + V\Bigg\}y\sqrt{z},\end{split}\ee
which is identical to the one (\ref{HD}) obtained following Dirac's constrained analysis. Action (\ref{AM1}) can again be written in the canonical (ADM) form \eqref{ADM} as before. Now, if all the coupling parameters $\alpha= \alpha_0$, $\beta = \beta_0$ and $\gamma= \gamma_0$ are constants, then $\alpha'=\beta'=\gamma'=0$ and so $M=0$. Additionally, if $k=0$, in that case the Hamiltonian takes the following form,

\be\label{HMconstant}\begin{split} &\mathcal{H_M} =  xp_z + wp_y + \frac{p_\phi^2}{2y\sqrt{z}} + \frac{1}{2(3\beta_0 + \gamma_0)}\Bigg[\left(8\beta_0 + 3\gamma_0\right)\frac{z^{\frac{3}{2}}p_x^2}{2y\gamma_0} + \left(2\beta_0 + \gamma_0\right)\frac{yp_w^2}{2\sqrt{z}\gamma_0}\\
&~~ - \frac{4\beta_0 + \gamma_0}{\gamma_0}\sqrt{z}p_xp_w - \frac{zp_x}{2}\left\{\left(4\beta_0 + \gamma_0\right)^2\frac{x^2}{2z^2\gamma_0} + \left(4\beta_0 + \gamma_0\right)\frac{xw}{zy} - \left(2\beta_0 + \gamma_0\right)\left(8\beta_0 + 3\gamma_0\right)\frac{w^2}{2y^2\gamma_0}\right\}\\
&~~ + \frac{yp_w}{2}\left\{ - \beta_0\frac{xw}{yz} + \left(2\beta_0 + \gamma_0)(4\beta_0 + \gamma_0\right)\frac{1}{2\gamma_0}\left(\frac{x^2}{z^2} - \frac{w^2}{y^2}\right)\right\}\Bigg] + \Bigg[\frac{1}{3\beta_0 + \gamma_0}\Bigg\{2\frac{\beta_0^3}{\gamma_0}\left(\frac{x^2}{z^2} - \frac{w^2}{y^2}\right)^2\\
&~~ + \beta_0^2\left(\frac{15x^4}{4z^4} + \frac{x^3w}{yz^3} - \frac{11x^2w^2}{2y^2z^2} + 6\frac{w^4}{y^4}\right) + \beta_0\gamma_0\left(\frac{25x^4}{12z^4} + \frac{2x^3w}{3yz^3} - \frac{7x^2w^2}{2y^2z^2} - \frac{4xw^3}{3y^3z} + \frac{43w^4}{12y^4}\right)\\
&~~ + \frac{\gamma_0^2}{4}\left(\frac{17x^4}{12z^4} + \frac{x^3w}{3yz^3} - \frac{5x^2w^2}{2y^2z^2} - \frac{5xw^3}{3y^3z}  + \frac{29w^4}{12y^4}\right)\Bigg\} + \alpha_0\Big(\frac{w^2}{2y^2} + \frac{wx}{yz}\Big) + V\Bigg]y\sqrt{z},\end{split}\ee

\section{Concluding remarks:}

Different canonical formalisms lead to different Hamiltonians for higher-order theories of gravity. These Hamiltonians although are canonically equivalent, lead to completely different quantum dynamics \cite{Ab9}. Due to nonlinearity, quantum canonical transformation is not possible. Following a series of works, it has been established that the divergent terms although do not affect classical phase-space structures, due to canonical transformation, these indeed tell upon the quantum dynamics, since as mentioned, quantum canonical transformation does not exist for non-linear theories such as gravity. In this sense, Modified Horowitz' formalism renders the correct quantum description. In fact, if Dirac's constrained analysis is followed only after taking care of the divergent terms, identical Hamiltonian is produced. Nevertheless, such canonical equivalence of higher-order theories of gravity between Dirac's constraint analysis, after taking care of the divergent terms appearing in  the action, and Modified Horowitz' formalism had been only established in the background of isotropic and homogeneous models. Here, we extend our work in anisotropic models to establish the same. We don't proceed any further towards canonical quantization, since this will turn out to be rather cumbersome due to operator ordering ambiguities between several pairs of operators in \eqref{HM2}, and at least between two pairs $\{x, p_x\}$ and $\{w, p_w\}$ in \eqref{HMconstant}. We believe that under suitable choice of coordinates, the quantum equation might turn out to be a bit simpler, which we pose in the future.

\appendix
\section{Field equations from the Hamiltonian:}

In order to justify the Hamiltonian $\mathcal{H_D}$ \eqref{HD}, in this appendix, we find the field equations in view of the Hamilton equations. Here, we need to find the form of $p_z$ and $p_y$. So, from Hamilton's equation we have

\be\label{pxdot}\begin{split} &\dot {p_x} = - \frac{\partial \mathcal{H_D}}{\partial x} = - \frac{1}{2(3\beta + \gamma)}\Bigg\{ - p_x(4\beta + \gamma)\Bigg((4\beta + \gamma)\frac{x}{z\gamma} + \frac{w}{y}\Bigg) + \frac{yp_w}{z}\Bigg((2\beta + \gamma)(4\beta + \gamma)\frac{x}{z\gamma}\\
&\hspace{0.5cm} - \frac{w\beta}{y}\Bigg)\Bigg\} - p_z  - m \frac{p_\phi}{z} - \Bigg\{mM  + \alpha \frac{w}{y} + \frac{1}{3\beta + \gamma}\Bigg(8\frac{\beta^3x}{\gamma z}\Big(\frac{x^2}{z^2} - \frac{w^2}{y^2}\Big) + \beta^2\Big(15\frac{x^3}{z^3} + 3\frac{wx^2}{yz^2}\\
&\hspace{0.5cm} - 11\frac{w^2x}{y^2z}\Big) + \beta\gamma\Big(\frac{25x^3}{3z^3} + 2\frac{wx^2}{yz^2} - 7\frac{w^2x}{y^2z} - \frac{4w^3}{3y^3}\Big) + \frac{\gamma^2}{4}\Big(\frac{17x^3}{3z^3} + \frac{wx^2}{yz^2} - 5\frac{w^2x}{y^2z} - \frac{5w^3}{3y^3}\Big)\Bigg)\Bigg\}\frac{y}{\sqrt{z}}\\
& \dot {p_w} = - \frac{\partial \mathcal{H_D}}{\partial w} = - p_y - n \frac{p_\phi}{y} - \frac{1}{2(3\beta + \gamma)}\Bigg\{ - \frac{p_xz}{y}\Bigg((4\beta + \gamma)\frac{x}{z} + (2\beta + \gamma)(8\beta + \gamma)\frac{w}{y\gamma}\Bigg)\\
&\hspace{1cm} - p_w\Bigg((2\beta + \gamma)(4\beta + \gamma)\frac{w}{y\gamma} + \frac{x\beta}{z}\Bigg)\Bigg\} - \Bigg\{nM  + \alpha\Bigg(\frac{x}{z}  + \frac{w}{y}\Bigg) + \frac{1}{3\beta + \gamma}\Bigg(\beta^2\Big(\frac{x^3}{z^3}\\
&\hspace{1cm} - 11\frac{wx^2}{yz^2} + 24\frac{w^3}{y^3}\Big) - 8\frac{\beta^3w}{y\gamma }\Big(\frac{x^2}{z^2} - \frac{w^2}{y^2}\Big) + \beta\gamma\Big(\frac{2x^3}{3z^3} - 7\frac{wx^2}{2yz^2} - 4\frac{w^2x}{y^2z} + \frac{43w^3}{3y^3}\Big)\\
&\hspace{1.0cm} + \frac{\gamma^2}{4}\Big(\frac{x^3}{3z^3} - 5\frac{wx^2}{yz^2} - 5\frac{w^2x}{y^2z} + \frac{29w^3}{3y^3} \Big)\Bigg) - 4k(3\beta + \gamma)\frac{w}{y^2}\Bigg\}\sqrt{z}\end{split}\ee
where, $m = \alpha' + 4\beta'\frac{k}{y} - (2\beta' + \gamma') \frac{x^2}{z^2}$ and $ n = 2\alpha' - (2\beta' + \gamma') \frac{w^2}{y^2} + 2(4\beta' + \gamma')\frac{k}{y}$. Now under the choice $N=1$ and using the expressions (\ref{momenta}) of $p_x$, $p_w$ and $p_\phi$, we can find the form of $p_z$ and $p_y$ from equations (\ref{pxdot}) as

\be\begin{split}& p_z = - \Bigg[\beta\Bigg(2\frac{\ddot x}{z} + 4\frac{\ddot w}{y} - 3\frac{\dot x\dot z}{z^2} - 2\frac{\dot w\dot z}{yz} - 3\frac{wx\dot z}{yz^2} + \frac{w^2\dot z}{2y^2z} + 2\frac{\dot x\dot y}{yz} + \frac{w^2\dot y}{y^3} + 6\frac{x\dot w}{yz} - 6\frac{w\dot w}{y^2} + \frac{7x^3}{3z^3} + 2\frac{wx^2}{yz^2}\\
&\hspace{1cm} - 3\frac{w^2x}{y^2z} + \frac{2w^3}{3y^3}\Bigg) + \gamma\Bigg(\frac{\ddot x}{z} + \frac{\ddot w}{y} - \frac{3\dot x\dot z}{2z^2} - \frac{\dot w\dot z}{2yz} - \frac{3wx\dot z}{4yz^2} + \frac{w^2\dot z}{4y^2z} + \frac{\dot x\dot y}{yz} + \frac{w^2\dot y}{2y^3} + \frac{3x\dot w}{2yz} - \frac{3w\dot w}{2y^2}\\
&\hspace{1cm} + \frac{7x^3}{6z^3} + \frac{wx^2}{4yz^2} - \frac{5w^2x}{4y^2z} - \frac{w^3}{6y^3}\Bigg) + \alpha\frac{w}{y} + \alpha'\dot\phi + \beta'\dot\phi\Bigg(2\frac{\dot x}{z} + 4\frac{\dot w}{y} - \frac{x^2}{z^2} + 2\frac{wx}{yz} - \frac{w^2}{y^2} + 4\frac{k}{y}\Bigg)\\
&\hspace{1cm} + \gamma'\dot\phi\Bigg(\frac{\dot x}{z} + \frac{\dot w}{y} - \frac{x^2}{2z^2}+ \frac{wx}{2yz} - \frac{w^2}{2y^2}\Bigg)\Bigg]\frac{y}{\sqrt{z}}\\
& p_y = - \Bigg[\alpha\Bigg(\frac{x}{z} + \frac{w}{y}\Bigg) + \beta\Bigg(4\frac{\ddot x}{z} + 8\frac{\ddot w}{y} - 2\frac{\dot x\dot z}{z^2} + 4\frac{\dot w\dot z}{yz} - 2\frac{wx\dot z}{yz^2} + 3\frac{x^2\dot z}{z^3} - 8\frac{\dot w\dot y}{y^2} - 4\frac{wx\dot y}{y^2z} + 6\frac{w\dot x}{yz}\\
&\hspace{1cm} - 6\frac{x\dot x}{z^2} + \frac{2x^3}{3z^3} - 3\frac{wx^2}{yz^2} + 2\frac{w^2x}{y^2z} + \frac{13w^3}{3y^3} - 12\frac{kw}{y^2}\Bigg) + \gamma\Bigg(\frac{\ddot x}{z} + 3\frac{\ddot w}{y} - \frac{\dot x\dot z}{2z^2} + \frac{3\dot w\dot z}{2yz} - \frac{wx\dot z}{4yz^2}\\
&\hspace{1cm} + \frac{3x^2\dot z}{4z^3} - 3\frac{\dot w\dot y}{y^2} - \frac{wx\dot y}{2y^2z} + \frac{3w\dot x}{2yz} - \frac{3x\dot x}{2z^2} + \frac{x^3}{12z^3} - \frac{5wx^2}{4yz^2} - \frac{w^2x}{2y^2z} + \frac{5w^3}{3y^3} - 4\frac{kw}{y^2}\Bigg) + 2\alpha'\dot\phi\\
&\hspace{0.8cm} + 2\beta'\dot\phi\Bigg(2\frac{\dot x}{z} + 4\frac{\dot w}{y} - \frac{x^2}{z^2} + 2\frac{wx}{yz} - \frac{w^2}{y^2} + 4\frac{k}{y}\Bigg) + \gamma'\dot\phi\Bigg(\frac{\dot x}{z} + 3\frac{\dot w}{y} - \frac{x^2}{2z^2} + \frac{wx}{2yz} - \frac{w^2}{y^2} + 2\frac{k}{y}\Bigg)\Bigg]\sqrt{z}\end{split}\ee
Knowing exact form of all momenta and by putting $x = \dot z$, $w = \dot y$, $z = A^2$, $y = B^2$, we are now able to write the $(^0_0)$ equation of Einstein as presented in (\ref{00}). Also, using the expressions $\lambda = p_z$ and $\tau = p_y$, from Lagrangian (\ref{L}), we can find the other field equations, viz., $A$, $B$, and $\phi$ variation equations as presented in equations (\ref{z}), (\ref{y}) and (\ref{phi}).

\end{document}